\pdfoutput=1
\documentclass{elsarticle}
\usepackage[utf8]{inputenc}

\usepackage{amsmath, amssymb}
\usepackage{booktabs}

\usepackage{tikz}
\usepackage{fp}

\newcommand{\rhoSI}{\rho_{\text{\sc si}}}
\newcommand{\rhoDSA}{\rho_{\text{\sc dsa}}}

\newcommand{\phiSI}{\phi_{\text{\sc si}}}
\newcommand{\phiDSA}{\phi_{\text{\sc dsa}}}

\newcommand{\phiDD}{\phi_{\text{\sc pdsa}}}
\newcommand{\rhoDD}{\widetilde\rho_{\text{d}}}
\newcommand{\rhoDDmax}{\rho_{\text{\sc pdsa}}^{\text{max}}}

\renewcommand{\vec}[1]{\mathbf{#1}}

\newcommand{\sigs}{\Sigma_{\text{s}}}

\newcommand{\err}[2]{e_{\text{\sc #1}}^{\text{#2}}}
\newcommand{\eNNl}{\err{nn}{l}}
\newcommand{\eNNr}{\err{nn}{r}}
\newcommand{\eDDl}{\err{dd}{l}}
\newcommand{\eDDr}{\err{dd}{r}}
\newcommand{\eDNl}{\err{dn}{l}}
\newcommand{\eNDr}{\err{nd}{r}}

\definecolor{color1}{HTML}{377EB8}
\definecolor{color2}{HTML}{E41A1C}
\definecolor{color3}{HTML}{4DAF4A}
\definecolor{color4}{HTML}{984EA3}
\definecolor{color5}{HTML}{FF7F00}
\definecolor{color6}{HTML}{A65628}
\definecolor{color7}{HTML}{F781BF}
\definecolor{color8}{HTML}{FFFF33}

\begin{document}

\begin{frontmatter}
  \title{Piecewise Diffusion Synthetic Acceleration Scheme for
    Neutron Transport Simulations in Diffusive Media}
  \author[edf]{François Févotte\corref{cor}}
  \ead{francois.fevotte@edf.fr}
  
  \address[edf]{EDF Lab Paris-Saclay -- 7, Boulevard Gaspard Monge -- 91120 Palaiseau, France}
  \cortext[cor]{Corresponding author}

  \begin{abstract}
    The method of discrete ordinates ($S_N$) is a popular choice for the
    solution of the neutron transport equation. It is however well known that it
    suffers from slow convergence of the scattering source in optically thick
    and diffusive media, such as pressurized water nuclear reactors (PWR). In
    reactor physics applications, the $S_N$ method is thus often accompanied by
    an acceleration algorithm, such as the Diffusion Synthetic Acceleration
    (DSA). With the recent increase in computational power, whole core transport
    calculations have become a reasonable objective. It however requires using
    large computers and parallelizing the transport solver. Due to the elliptic
    nature of the DSA operator, its parallelization is not straightforward. In
    this paper, we present an acceleration operator derived from the DSA, but
    defined in a piecewise way such that its parallel implementation is
    straightforward. We mathematically show that, for optically thick enough
    media, this Piecewise Diffusion Synthetic Acceleration (PDSA) preserves the
    good properties of the DSA. This conclusion is supported by numerical
    experiments.
  \end{abstract}

\end{frontmatter}

\section{Introduction}

The simulation of neutron transport phenomena in nuclear reactor cores requires
the solution of the Boltzmann Transport Equation~(BTE). We focus in this paper
on the simulation of Pressurized Water Reactors~(PWR). For such reactors, the
geometry and materials used make the domain optically thick and diffusive,
meaning \textit{(i)}~that the core size represents a large number of neutron
mean free paths, and \textit{(ii)}~that scattering represents a large fraction
of neutron-matter interactions. In such cases, the diffusion equation is often
considered a good enough alternative to the BTE, which is why most industrial
calculations rely on a 2-step scheme:
\begin{enumerate}
\item the BTE is solved in 2D, at the scale of an assembly, with relatively fine
  spatial and energetic discretization. This calculation produces homogenized
  and condensed cross-sections;
\item these homogenized and condensed sections are fed to a 3D diffusion or
  Simplified Transport~($SP_{N}$) calculation, which is performed at the scale
  of the reactor core and uses a relatively coarse spatial and energetic
  discretization.
\end{enumerate}

Such a scheme presents the advantage of involving neutron transport calculations
only at the scale of the fuel assembly and for two spatial dimensions. In such
calculations, energy is traditionally discretized using the multigroup
formalism, and the angular variable is handled by the discrete ordinates~($S_N$)
method. Various methods can be used to discretize the spatial operators, but we
will not enter such details in this paper. As was uncovered by early adopters of
the $S_N$ formalism, this method suffers from a major problem in optically thick
diffusive media: the classical Source Iterations~(SI) converge very slowly in
this case. To remedy this issue, the Diffusion Synthetic Acceleration~(DSA)
scheme has been proposed as early as the late
1970s~\cite{alcouffe1977,larsen1984,adams2002}, and probably remains one of the
most popular acceleration schemes today, especially for Cartesian geometries.

\bigskip

However, the approximations induced by the use of such 2-step schemes need to be
assessed, which is why full core 3D neutron transport solvers are still
needed. We focus here on the solution of the 3D stationary BTE, which is one of
the most important building blocks for state-of-the-art 3D whole-core
criticality calculations. Even though it ignores the time variable, the 3D
stationary BTE is still set in a 6-dimensional phase space (3 for space, 2 for
travel direction and 1 for energy). Its discretization at the scale of the full
reactor core therefore quickly produces very large problems of size in the order
of $10^{10}$ to $10^{12}$ degrees of freedom, whose solution has remained mainly
out of reach before the early 2010s~\cite{davidson2011,Cou12,courau2013}, when
large enough supercomputers became available, along with numerical methods able
to efficiently harness them.

Devising and implementing parallel methods able to efficiently solve the
transport equation for such large problems is in itself no easy task, the major
difficulty lying in the fact that the hyperbolic nature of the transport
equations implies dependencies between cells. However, another practical
difficulty arises, in the case of optically thick geometries, from the need for
an acceleration scheme that \textit{(i)}~accelerates Source Iterations, and
\textit{(ii)}~can be efficiently parallelized using the same data distribution
as the transport solver.

A first technique consists in keeping the traditional DSA scheme, and
parallelizing it alongside the transport solver. This presents the advantage of
reusing the same whole-core diffusion solvers as the second step mentioned
above. However, industry-grade neutron diffusion solvers are generally
sequential, and the elliptic nature of the diffusion equation makes their
parallelization a challenging task. Although efficient parallel diffusion
solvers can be implemented~\cite{barrault2011,jamelot2013}, the induced code
complexity is often considered a heavy price to pay. The same is also true in
the case of alternate acceleration methods such as Coarse-Mesh Finite
Differences~\cite{lenain2014}, which are also elliptic in nature and thus
difficult to parallelize.

Other techniques consist in departing from the standard Source Iterations + DSA
scheme. For example, DENOVO uses a Krylov solver~\cite{davidson2011}, which
converges faster than the traditional multi-group Gauss-Seidel algorithm and
angular Source Iterations and alleviates the need for an acceleration
scheme. Such a Krylov solver can still be further preconditioned, for example
using multigrid methods in energy~\cite{slaybough2013}. While very efficient,
the implementation of such techniques makes the reference neutron transport code
share few software components, or even algorithms, with the industrial diffusion
code. This, once again, makes the development, maintenance and verification
price heavy to pay for the industry.

\bigskip

In this paper, we introduce the Piecewise Diffusion Synthetic Acceleration
scheme (PDSA), a new acceleration method for parallel neutron transport
calculations, specifically designed to minimize the development effort and reuse
as much as possible existing diffusion solvers. Indeed, the scheme is defined in
such a way that any code implementing transport iterations accelerated by a DSA
operator (with consistent spatial discretization schemes), can be transformed in
a PDSA implementation at practically no programming cost. We will focus here on
the definition of the PDSA scheme, and on the proof that it converges at the
continuous level, along with simple 1D numerical experiments. We show in a
companion paper~\cite{moustafa20xx} how this has been implemented in EDF's
COCAGNE~\cite{calloo2017} platform, which features a diamond-difference $S_N$
transport solver, accelerated by an $SP_1$ solver using mixed dual
Raviart-Thomas~($RT_k$) finite elements~\cite{hebert2006,courau2013}.

The remainder of this paper is organized as follows: in the following part, we
briefly describe the PDSA scheme. We then proceed to a Fourier analysis in
part~\ref{sec:fourier}: we review the standard unaccelerated transport source
iterations, as well as the DSA scheme. Then, we Fourier analyze the proposed
PDSA scheme. We show in particular how it can be seen as a perturbation of the
standard DSA scheme, and derive conditions under which the perturbation is small
enough that convergence properties of the DSA are not lost. In
part~\ref{sec:num}, we assess the validity of the theory by performing a few
numerical experiments in 1D. We finally make a few concluding notes in
part~\ref{sec:concl}.

\section{Description of the Piecewise Diffusion Synthetic Acceleration}
\label{sec:pdsa}

In this section, we briefly describe and introduce the Piecewise Diffusion
Synthetic Acceleration scheme. The focus here is on the definition of the
scheme, while part~\ref{sec:fourier.pdsa} will be devoted to the analysis of its
properties.

\subsection{Standard DSA}

As underlined in the introduction, the PDSA is defined as a perturbation to the
DSA scheme. We thus start with recalling the standard equations of the DSA. Let
us consider the following time-independent, one-group neutron transport equation
with isotropic scattering:
\begin{align}
  \label{eq:md.trans}
  &\forall \vec\Omega \in S^2, \forall\vec{r}\in{\cal D},\notag\\
  &\qquad\vec\Omega \cdot \vec\nabla\psi(\vec{r},\vec\Omega)
    + \Sigma(\vec{r})\,\psi(\vec{r},\vec\Omega)
    = Q(\vec{r}) + \frac{\sigs(\vec{r})}{4\,\pi}
    \int_{S_2} d\vec\Omega^\prime \;
    \psi(\vec{r},\vec\Omega^\prime),
\end{align}
with void boundary conditions to model a full core:
\begin{align}\label{eq:md.trans.bc}
  \begin{aligned}
    &\forall\vec{r}\in\partial{\cal D}, \forall \vec\Omega \in S^2
    \text{ such that }\vec\Omega\cdot\vec{n}(\vec{r}) < 0, \\
    &\qquad \psi(\vec{r},\vec\Omega) = 0.
  \end{aligned}
\end{align}
In the equation above, $\psi(\vec{r},\vec\Omega)$ denotes the neutron flux at
position~$\vec{r}$ and in direction~$\vec\Omega$. The total and scattering
cross-sections are denoted by $\Sigma$ and $\sigs$ respectively, and $Q$ is a
source term coming from outer iterations. The spatial domain is denoted by
${\cal D}$, and its boundary by $\partial{\cal D}$. The normal vector to this
boundary is $\vec{n}$, so that the boundary condition above states that no flux
enters the domain.

In this context, the traditional DSA scheme is defined as follows. In a first
stage, the streaming operator is inversed:
\begin{align}\label{eq:md.dsa.trans}
  \vec\Omega \cdot \vec\nabla\psi_{\ell+\frac{1}{2}}(\vec{r},\vec\Omega) +
  \Sigma(\vec{r})\,\psi_{\ell+\frac{1}{2}}(\vec{r},\vec\Omega)
  = Q(\vec{r}) + \sigs(\vec{r}) \; \phi_\ell(\vec{r}),
\end{align}
where inner iteration index~$\ell$ was introduced, along with the scalar flux
\begin{align*}
  \phi_\ell(\vec{r}) = \frac{1}{4\,\pi} \int_{S_2} d\vec\Omega \; \psi_\ell(\vec{r},\vec\Omega).
\end{align*}
In a second stage, an approximate diffusion operator is solved
\begin{align}\label{eq:md.dsa.diff}
  \text{div}\left(\frac{1}{3\,\Sigma}\nabla\widetilde\phi_{\ell+1}\right)
  + \Sigma\,\widetilde\phi_{\ell+1}
  = \sigs \;
  \left(\phi_{\ell+\frac{1}{2}} - \phi_{l}\right).
\end{align}

While the original void boundary conditions~\eqref{eq:md.trans.bc} can be retained
for transport equation~\eqref{eq:md.dsa.trans}, they have no meaning for
diffusion equation~\eqref{eq:md.dsa.diff} whose unknown is a scalar flux. They
are thus usually replaced by homogeneous Dirichlet boundary conditions:
\begin{align}\label{eq:md.dsa.diff.bc}
\widetilde\phi(\vec{r}) = 0, \qquad\forall\vec{r}\in\partial{\cal D}.
\end{align}

At the end of a Diffusion Synthetic Accelerated iteration, the scalar flux is updated as follows:
\begin{align*}
  \phi_{\ell+1} = \phi_{\ell+\frac{1}{2}} + \widetilde\phi_{\ell+1}.
\end{align*}

\subsection{Piecewise DSA}

\begin{figure}[tb]
  \begin{center}
    \includegraphics{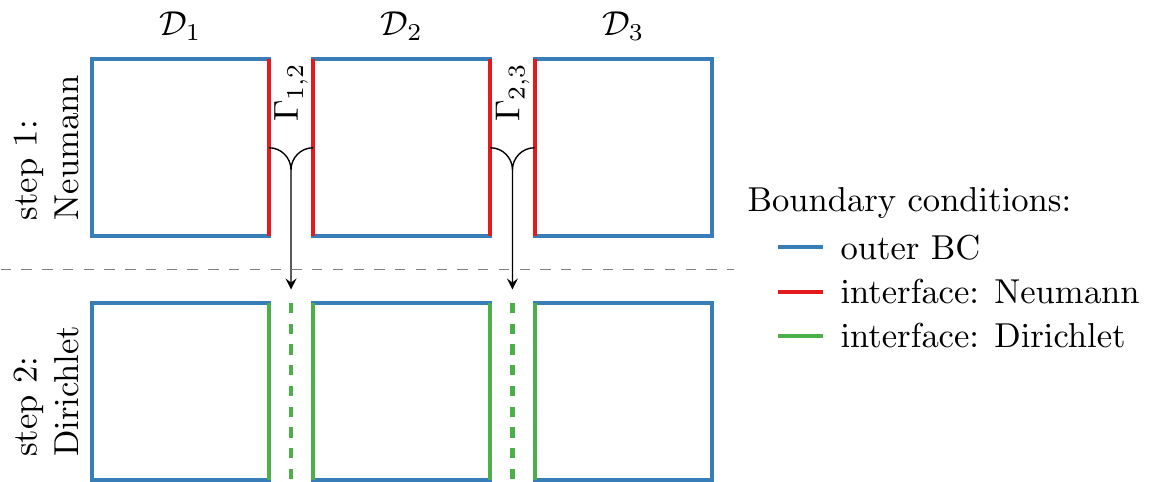}
  \end{center}
  \caption{Schematic presentation of the PDSA scheme in a 3-subdomain case}
  \label{fig:pdsa}
\end{figure}

The PDSA scheme described in this paper aims at replacing
system~\mbox{\eqref{eq:md.dsa.diff}--\eqref{eq:md.dsa.diff.bc}} by an operator
which is more local and easier to solve in parallel. Figure~\ref{fig:pdsa}
illustrates the construction of PDSA on domain ${\cal D}$, which has been
partitioned into
\begin{align*}
  {\cal D} = \cup_{i=1}^N {\cal D}_i,
\end{align*}
with $N=3$ in the figure. In the following, we temporarily drop iteration
indices~$\ell$ to simplify the notations.

\newcommand{\phiN}{\widetilde\phi_{\text{\sc n}}}
\newcommand{\phiD}[1]{\widetilde\phi_{{\text{\sc d}}#1}}

In a first step, called the Neumann diffusion problem in the following, a flux
correction~$\phiN^i$ is computed as the solution to
equation~\eqref{eq:md.dsa.diff} in each subdomain~${\cal D}_i$. Boundary
condition~\eqref{eq:md.dsa.diff.bc} is considered for the outer
boundary~$\partial{\cal D}\cap\partial{\cal D}_i$. However, at inner interfaces
between subdomains, an homogeneous Neumann boundary condition is used:
\begin{align}\label{eq:md.pdsa.neumann}
  \nabla\phiN^i(\vec r) \cdot \vec{n}(\vec{r}) = 0,
  \qquad\forall\vec{r}\in\Gamma_i = \partial{\cal D}_i \setminus \partial{\cal D}.
\end{align}

The second step, hereafter called the Dirichlet diffusion problem, differs from
the first only with respect to the boundary conditions at the interface. A flux
correction~$\phiD{}^i$ is computed as the solution to
equation~\eqref{eq:md.dsa.diff} in each subdomain, but in this case an
inhomogeneous Dirichlet boundary condition is used at the interface between
subdomains: for any two subdomains ${\cal D}_i$ and ${\cal D}_j$ sharing a
common interface $\Gamma_{i,j}$,
\newcommand{\trace}{\text{\Large$\gamma$}_{\Gamma_{i,j}}}
\begin{align*}
  \phiD{}^i(\vec{r}) = \phiD{}^j(\vec{r}) = \frac{1}{2} \left( \trace(\phiN^i) + \trace(\phiN^j)\right),
  \qquad\forall\vec{r}\in\Gamma_{i,j}.
\end{align*}
In the equation above, $\trace$ denotes the trace function on the $\Gamma$
interface, so that the value at the interface is computed as the half-sum of
values coming from both subdomains at first step.

\medskip

The solution to this second step is used to update the scalar flux at the end of
a PDSA iteration:
\begin{align*}
  \phi_{\ell+1}(\vec{r}) = \phi_{\ell+\frac{1}{2}}(\vec{r}) +
  \phiD{,\ell+1}(\vec{r}),
  \qquad\forall\vec{r}\in{\cal D}_i, \quad \forall i.
\end{align*}

\subsection{Advantages and limits of PDSA}

As will be shown by a Fourier analysis in the next section, under some
circumstances (when subdomains are optically thick enough), the PDSA operator
can accelerate source iterations. When this is the case, it features some
advantages over the standard DSA scheme.

First and foremost, it is defined piecewise, which means it can adapt to any
geometric domain partitioning used by the underlying transport solver. In a
parallel context, there is only one point-to-point data exchange, so that the
communication overhead can be considered low with respect to the computations
performed within each subdomain.

Also, the two steps of PDSA are very similar problems. If an iterative solver is
used to solve them, the second step can be initialized with the first to help it
converge faster.

Moreover, if a DSA scheme is already available in a given neutron transport code
(\textit{i.e.} if a diffusion solver has already been developed, with a
consistent discretization scheme), PDSA can be implemented at almost no
additional cost\footnote{The only missing feature might be inhomogeneous Dirichlet
boundary conditions, which are not always implemented in diffusion solvers.}. This
allows for easy parallelization of the acceleration scheme when parallelizing
the transport solver.

\medskip

Limitations of the PDSA scheme obviously lie in the conditions under which it
accelerates the source iterations. This limits the number of subdomains which can
be defined for a given calculation. It should however be noted that
section~\ref{sec:fourier.pdsa} gives indicators which can be computed beforehand
to estimate the maximal number of subdomains allowed, or warn a user if the
computation might not converge. Although this does not alleviate the limitation
in the number of subdomains, it at least allows avoiding most common mistakes.

\subsection{Relationship to Domain Decomposition Methods}

It should be noted that the two diffusion steps (Neumann and Dirichlet) in PDSA
correspond to the first iteration of a Domain Decomposition~(DD) technique
called the Dirichlet--Dirichlet algorithm in~\cite{toselli2000}, or the
Dirichlet preconditioned FETI method introduced in~\cite{farhat1994}. It is also
related to Neumann--Neumann methods (which have been studied as early
as~\cite{glowinski1987}), in which each iteration defines the same two steps in
the reverse order (the Dirichlet problem is solved first, and imposes a boundary
condition to the Neumann problem).

All these domain decomposition techniques differ from the PDSA scheme proposed
here, in that they are defined as iterative methods, \textit{i.e.} it is proved
that, whatever the value imposed on the interface at the first iteration, they
converge to the solution of the diffusion problem when multiple iterations are
performed, but nothing is said of the solution given after the first
iteration. In our case, we impose a null current boundary condition in the
Neumann step~\eqref{eq:md.pdsa.neumann}, and are able to prove that one
iteration is enough to make the PDSA scheme convergent under some
assumptions.

Should these conditions become too restrictive in practice, then a potential
solution could be to add more Dirichlet--Dirichlet iterations. and turn PDSA
into a full domain-decomposition technique. However, this increases the number
of computations (and the number of communications in a parallel setup). In such
a case, it would be interesting to compare the efficiency of the
Dirichlet--Dirichlet method to other DD techniques such as the one described
in~\cite{jamelot2013}. Such a comparison should be performed in the specific
case of DSA problems, since the diffusion solver is only required to attenuate
some error modes in this context.

\section{Fourier Analysis}
\label{sec:fourier}

The 1D Fourier analysis is the primary tool used in the literature for the study
of acceleration schemes~\cite{adams2002}. In this section, we briefly review the
well-known Fourier analysis of the standard source iterations and DSA schemes,
before extending it to the proposed PDSA scheme.

We will perform this analysis on the case of an homogeneous infinite 1D slab
geometry, modeled by the finite spatial domain ${\cal D} = [0, L]$, with
reflective boundary conditions. In this case, the neutron transport problem can
be written as:
\begin{align}\label{eq:transport}
  \left\{
    \begin{aligned}
      &\mu \frac{\partial\psi}{\partial x}(x,\mu) + \Sigma\,\psi(x,\mu)
      = \frac{\sigs}{2}\,\int_{-1}^1 \psi(x, \mu') \; d\mu' + Q(x), \\[0.5em]
      &\psi(0,\mu) = \psi(0, -\mu),\\[0.5em]
      &\psi(L,\mu) = \psi(L, -\mu),\\
    \end{aligned}
  \right.
\end{align}
where notations are consistent with equation~\eqref{eq:md.trans}, except that in
a 1D geometry, $x$ represents the spatial variable and $\mu$ is the cosine of
the angular direction. In the following, the scattering ratio will be denoted by
$c=\frac{\sigs}{\Sigma}$. It will be assumed to be strictly less than 1, in
order for the transport problem to be well posed.

\subsection{Source Iterations}

The standard source iterations scheme is defined as:
\begin{align}\label{eq:SI}
  \left\{
    \begin{aligned}
      &\mu \frac{\partial\psi_{\ell+1}}{\partial x}(x,\mu) + \Sigma\,\psi_{\ell+1}(x,\mu)
      = c\,\Sigma\,\phi_{\ell} + Q(x), \\[0.5em]
      &\psi_{\ell+1}(0,\mu) = \psi_{\ell+1}(0, -\mu),\\[0.5em]
      &\psi_{\ell+1}(L,\mu) = \psi_{\ell+1}(L, -\mu),\\
    \end{aligned}
  \right.
\end{align}
in which the scalar flux was introduced:
\begin{align}
  \phi_{\ell}(x) &= \frac{1}{2}\,\int_{-1}^1 \psi_{\ell}(x, \mu') \; d\mu'.\notag
\end{align}

The error after the $\ell$\textsuperscript{th} iteration can be defined as
$e_{\ell} = \psi(x,\mu) - \psi_{\ell}(x,\mu)$. This error follows the same
scheme as~\eqref{eq:SI}, but with $Q(x) = 0$. Analyzing the convergence of the
source iterations scheme towards $\psi$ for an arbitrary $Q$ source term is thus
equivalent to studying the convergence towards $0$ without source term. In the
following, we will thus consider~$Q=0$ and consider the flux~$\psi_{\ell}$ to be an
error term~$e_{\ell}$ of which we will study the convergence towards 0.

The efficiency of the source iterations scheme is traditionally studied using a
Fourier analysis. Assuming the (scalar) initial error to be given by
\begin{align}
  \phi_0(x) &= \cos \left({{\pi\,k\,x}\over{L}}\right)
, \label{eq:phi0}
\end{align}
then the first iteration yields the angular flux
\begin{align*}
  \psi_1(x, \mu) &= {{c\,\mu\,\omega\,\sin \left(\omega\,\Sigma\,x\right)+c\,\cos \left(
 \omega\,\Sigma\,x\right)}\over{\mu^2\,\omega^2+1}}
,
\end{align*}
where $\omega={{\pi\,k}\over{\Sigma\,L}}
$ denotes the frequency of the initial
error. After the first source iteration, the scalar flux is given by
\begin{align*}
  \phiSI(x)
  &= \frac{1}{2} \int_{-1}^{1} \psi_1(x, \mu)\;d\mu\\
  &= \rhoSI(\omega) \, \phi_0(x),
\end{align*}
where subscript {\sc si} denotes the Source Iterations scheme and
\begin{align*}
  \rhoSI(\omega) = {{c\,\arctan \omega}\over{\omega}}
.
\end{align*}

In other words, functions of the form given by $\phi_0$ are eigenmodes of
the source iteration operator, associated to eigenvalues $\rhoSI$. After $\ell$
source iterations, the scalar error is given by the expression:
\begin{align*}
  \phi_{\text{\sc si},\ell}(x) &= \rhoSI^\ell(\omega) \; \phi_0(x).
\end{align*}

\begin{figure}[th]
  \begin{center}
    \includegraphics[width=\linewidth]{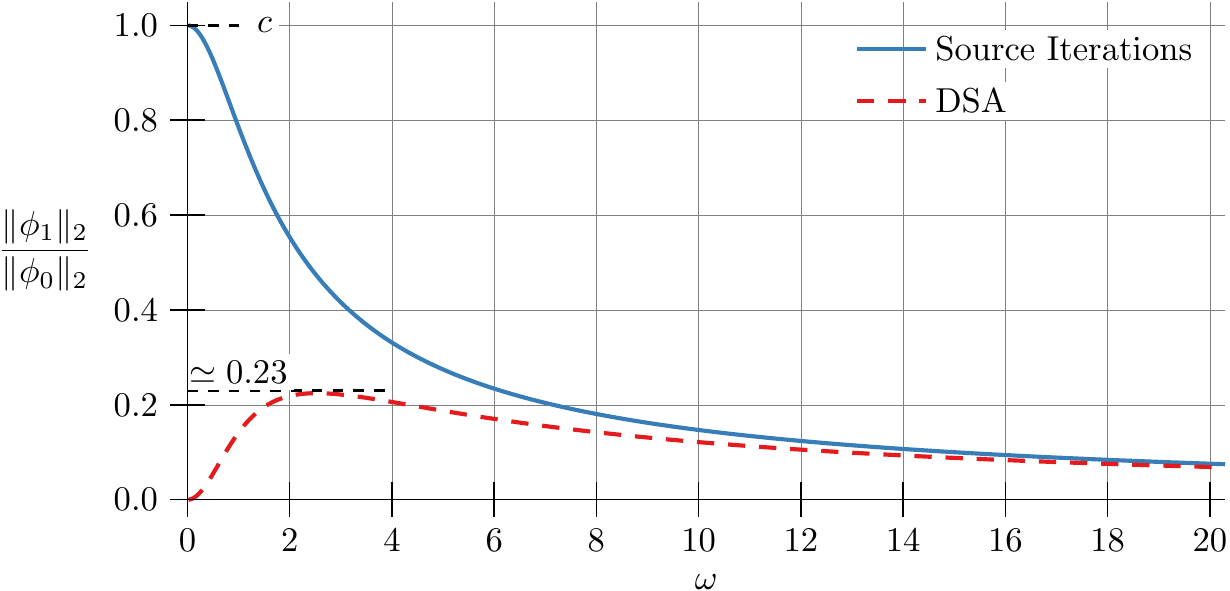}
  \end{center}
  \caption{Amplification factors of the Source Iteration and DSA schemes as functions of
    frequency~$\omega$, in the diffusive case ($c=1$).}
  \label{fig:rho}
\end{figure}

The solid line in figure~\ref{fig:rho} presents the evolution of~$\rhoSI$ as a
function of frequency~$\omega$. It shows that the spectral radius of the source
iteration scheme is \mbox{$c$}, obtained for~\mbox{$\omega = 0$}. In diffusive
media, convergence can thus become arbitrarily slow. The slowest modes are
defined by low frequencies \mbox{($\omega \ll 1$)}, and correspond to a weak
spatial and angular dependency:
\begin{align*}
  \psi_1(x,\mu) \mathop{~~\sim~~}_{\omega\rightarrow 0} c
 + O(\omega^2).
\end{align*}
This shows that the source iterations scheme needs to be accelerated, and that
the acceleration operator will be most effective if it allows correctly handling
slowly oscillating modes.

\subsection{Diffusion Synthetic Acceleration}

In this section, we describe the Diffusion Synthetic Acceleration scheme, which
can be used to improve the convergence properties of the Source Iterations scheme.

\subsubsection{Diffusion Problem}

The accelerated scheme starts by a standard source iteration on a transport
operator, as described in equation~\eqref{eq:SI}. Subtracting the source
iteration equation from the exact transport equation~\eqref{eq:transport} yields
equations verified by error
\begin{align}
F_\ell(x,\mu) &= \psi(x,\mu) - \psi_\ell(x,\mu).\notag
\end{align}
After rearrangement of the terms, and adding the boundary conditions, equations
followed by the error are given by:
\begin{align}
  \left\{
    \begin{aligned}
      &\mu \frac{\partial F_{\ell+1}}{\partial x}(x,\mu) + \Sigma\,F_{\ell+1}(x,\mu)
      = c\,\Sigma\,\left[\vphantom\sum \phi_{\ell+1}(x)-\phi_{\ell}(x)\right] , \\[0.5em]
      &F_{\ell+1}(0,\mu) = F_{\ell+1}(0, -\mu),\\[0.5em]
      &F_{\ell+1}(L,\mu) = F_{\ell+1}(L, -\mu).\\
    \end{aligned}
  \right.\notag
\end{align}

This problem is of course as complicated to solve as the initial transport
problem. The principle of the DSA scheme consists in replacing it with an
approximated diffusion problem, whose solution is easier to compute. At the
first iteration ($l=0$), one thus computes the solution to the following problem:
\begin{equation}\label{eq:diffusion.glob}
  \left\{
    \begin{aligned}
      &\frac{-1}{3\Sigma} \, f''(x) + (1-c)\,\Sigma\,f(x)
      = c\,\Sigma\,\left[\vphantom\sum \phiSI(x) - \phi_0(x)\right], \\[0.5em]
      &f'(0) = 0, \\[0.5em]
      &f'(L) = 0.
    \end{aligned}
  \right.
\end{equation}
In the problem above, unknown $f$ is supposed to be an approximation to the
scalar flux associated to error $F_1$:
\begin{align}
  f(x) \simeq \frac{1}{2} \, \int_{-1}^{1} F_1(x,\mu') \; d\mu'.\notag
\end{align}

We can show that the solution to this problem takes the form
\begin{align*}
  f(x) = \rho_d(\omega) \, \phi_0(x),
\end{align*}
where subscript $d$ denotes that it comes from a diffusion calculation, and we
introduced
\begin{align*}
  \rho_d(\omega) = {{c\,\left(3\,\rhoSI(\omega)-3\right)}\over{\omega^2-3\,c+3}}
.
\end{align*}

Once again, this shows that functions of the form given by $\phi_0$ are
eigenmodes of the diffusion operator.

\subsubsection{Flux correction}

At the end of a diffusion-accelerated iteration, the scalar flux is given by
\begin{align*}
  \phiDSA
  &= \phiSI + f\\
  &= (\rhoSI + \rho_d) \; \phi_0\\
  &= \rhoDSA \; \phi_0,
\end{align*}
where subscripts {\sc dsa} denote that the quantities are defined in the DSA
scheme, and the eigenvalue associated to $\phi_0$ for the whole iteration is
denoted by
\begin{align*}
  \rhoDSA(\omega) &= {{\omega^2\,\rhoSI(\omega)+3\,\rhoSI(\omega)-3\,c}\over{\omega^2-3\,c
 +3}}
.
\end{align*}

The dashed line of figure~\ref{fig:rho} presents, in the diffusive case~($c=1$), the variation
of~$\rhoDSA$ as a function of frequency~$\omega$. It shows that low frequency
modes (\mbox{$\omega\ll 1$}) are associated to significantly lower eigenvalues in the
DSA scheme than in the Source Iteration scheme. The spectral radius of the DSA
iteration is approximately~$0.23$, obtained for~$\omega\simeq 2.5$. This shows
that the DSA scheme presents much more interesting convergence properties than
the source iterations scheme (see for example~\cite{adams2002} for a more
thorough analysis of the DSA scheme).

\subsection{Piecewise Diffusion Synthetic Acceleration (PDSA)}
\label{sec:fourier.pdsa}

\newcommand{\point}[3]{%
  \draw (#2,-0.1) -- (#2,0.1) ;
  \draw (#2,-0.1) node(#1)[below]{#3}
}
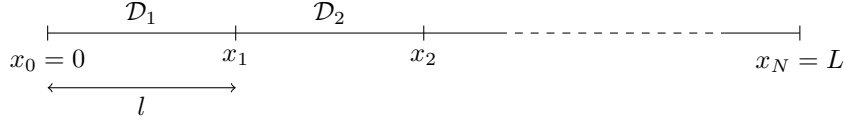
\begin{figure}[th]
  \centering
  \begin{tikzpicture}[scale=1]
    \draw (0,0) -- (6,0);
    \draw [dashed] (6,0) -- (9,0);
    \draw (9,0) -- (10,0);
    
    \point{A}{0}{$x_0 = 0$};
    \point{B}{2.5}{$x_1$};
    \point{C}{5}{$x_2$};
    \point{D}{10}{$x_N = L$};
    \draw (1.25, 0) node[above]{${\cal D}_1$};
    \draw (3.75, 0) node[above]{${\cal D}_2$};

    \draw [<->] (A.south) ++(0,-0.1) -- ++(2.5,0) node[midway,anchor=north]{$l$};
  \end{tikzpicture}
  \caption{Partition in subdomains for the Piecewise Diffusion Synthetic
    Acceleration (PDSA) scheme.}
  \label{fig:dd}
\end{figure}

We now perform the same analysis, replacing the standard DSA scheme by the PDSA
scheme introduced in section~\ref{sec:pdsa}.
Domain ${\cal D}$ is partitioned in $N$ subdomains without overlapping. In the
remaining of this paper, the following notations will be used, as explained on
figure~\ref{fig:dd}:
\begin{alignat*}{2}
  l   &= \frac{L}{N}, &\quad\\[0.5em]
  x_i &= i\;l, && 0 \leqslant i \leqslant N,\\[0.5em]
  {\cal D}_i &= [x_{i-1}, x_{i}], && 1 \leqslant i \leqslant N,\\[0.5em]
  t_i &: \begin{array}[t]{rcl}
           {\cal D}_i & \rightarrow & {\cal D}_1\\
           x & \mapsto & x-x_{i-1},
         \end{array} && 1 \leqslant i \leqslant N.
\end{alignat*}
Each subdomain is a segment of length~$l$. Translation~$t_i$ maps
subdomain~${\cal D}_i$ onto the reference subdomain~${\cal D}_1 = [0, l]$.

\medskip

The Piecewise Diffusion Synthetic Acceleration is defined by the following steps:
\begin{enumerate}
\item a transport source iteration~\eqref{eq:SI} is performed, yielding scalar
  flux $\phiSI$;
\item a diffusion problem is solved in each subdomain, with outer boundary
  conditions coming from~\eqref{eq:diffusion.glob}, and homogeneous Neumann
  conditions at interfaces between subdomains:
  \begin{equation}\label{eq:diffusion.nn}
    \left\{
      \begin{aligned}
        &\frac{-1}{3\Sigma} \, g^{\prime\prime}(x) + (1-c)\,\Sigma\,g(x)
        = c\,\Sigma\,\left[\vphantom\sum \phiSI(x) - \phi_0(x)\right], \\[0.5em]
        &g^\prime(0) = g^\prime(L) = 0, \\[0.5em]
        &g^\prime(x_i) = 0, \qquad 1 \leqslant i \leqslant N-1.
      \end{aligned}
    \right.
  \end{equation}
\item a second diffusion problem is solved in each subdomain, again with outer
  boundary conditions from~\eqref{eq:diffusion.glob}, but now with inhomoegenous
  Dirichlet conditions at the interfaces. The value set for the flux at the
  interfaces is obtained as the half sum of the interface values of the
  solutions in the previous step:
  \begin{equation}\label{eq:diffusion.dd}
    \left\{
      \begin{aligned}
        &\frac{-1}{3\,\Sigma} \, h^{\prime\prime}(x) + (1-c)\,\Sigma\,h(x)
        = c\,\Sigma\,\left[\vphantom\sum \phiSI(x) - \phi_0(x)\right], \\[0.5em]
        &h^\prime(0) = h^\prime(L) = 0, \\[0.5em]
        &h(x_i) = \frac{1}{2} \left[ g^-(x_i) + g^+(x_i) \right], \qquad 1 \leqslant i \leqslant N-1.
      \end{aligned}
    \right.
  \end{equation}
\item at the end of an iteration, the scalar flux is corrected using the
  solution of the second diffusion problem:
  \begin{align*}
    \phiDD &= \phiSI + h.
  \end{align*}
\end{enumerate}

\subsubsection{Step 1: Neumann Diffusion Problem}

We study here the PDSA scheme as a perturbation of the DSA scheme. We therefore
consider the error introduced by the Neumann Diffusion
problem~\eqref{eq:diffusion.nn}, with respect to the global diffusion
problem~\eqref{eq:diffusion.glob}:
\begin{align*}
  \delta &= g - f.
\end{align*}

Subtracting~\eqref{eq:diffusion.glob} to~\eqref{eq:diffusion.nn},
rearranging the terms, and restricting it to~${\cal D}_1$, we find
that~$\delta_{|{\cal D}_1}$ verifies
\begin{align*}
  \left\{
    \begin{aligned}
      &\frac{-1}{3\,\Sigma} \, \delta_{|{\cal D}_1}^{\prime\prime}(x) + (1-c)\,\Sigma\,\delta_{|{\cal D}_1}(x) = 0, \\[0.5em]
      &\delta_{|{\cal D}_1}^\prime(x_0) = -f^\prime(x_0), \\[0.5em]
      &\delta_{|{\cal D}_1}^\prime(x_1) = -f^\prime(x_1).
    \end{aligned}
  \right.
\end{align*}
$\delta_{|{\cal D}_1}$ can thus be defined as the linear combination
\begin{align*}
  \delta_{|{\cal D}_1} &= -f^\prime(x_0) \; \eNNl - f^\prime(x_1) \; \eNNr,
\end{align*}
where functions $\eNNl$ and  $\eNNr$ measure the error due to not knowing
the real boundary values of $f^\prime$ respectively on the left and right extremities
of ${\cal D}_1$:
\begin{align*}
  \left\{
    \begin{aligned}
      &\frac{d^2\eNNl}{dx^2}(x) - \alpha^2 \; \eNNl(x) = 0, \\[0.5em]
      &\frac{d\eNNl}{dx}(0) = 1, \\[0.5em]
      &\frac{d\eNNl}{dx}(l) = 0,
    \end{aligned}
  \right.
  &\qquad\text{ and }\qquad
  \left\{
    \begin{aligned}
      &\frac{d^2\eNNr}{dx^2}(x) - \alpha^2 \; \eNNr(x) = 0, \\[0.5em]
      &\frac{d\eNNr}{dx}(0) = 0, \\[0.5em]
      &\frac{d\eNNr}{dx}(l) = 1,
    \end{aligned}
  \right.
\end{align*}
where parameter $\alpha$ is defined as
\begin{align*}
  \alpha &= \sqrt{3\;(1-c)} \; \Sigma.
\end{align*}

Solutions to these problems can be analytically calculated, and are linked by
symmetry relations:
\begin{align}
  \eNNl(x) = -{{e^ {- \alpha\,x }\,\left(e^{2\,\alpha\,x}+e^{2\,\alpha\,l}\right)
 }\over{\alpha\,e^{2\,\alpha\,l}-\alpha}}

  \quad\text{and}\quad
  \eNNr(x) = -\eNNl(l-x).
  \label{eq:enn.sym}
\end{align}

Finally, any subdomain ${\cal D}_i$ can be mapped to ${\cal D}_1$ using
translation $t_i$, which allows following the same line of reasoning to obtain:
\begin{align}\label{eq:delta}
  \delta &= - \sum_{i=1}^{N}
     \chi_i \left[ f^\prime(x_{i-1}) \; \eNNl + f^\prime(x_{i}) \; \eNNr \right]
           \circ t_i,
\end{align}
where $\chi_i$ denotes the indicator function for subdomain ${\cal D}_i$.

\subsubsection{Step 2: Dirichlet Diffusion Problem}

We now consider the Dirichlet Diffusion problem~\eqref{eq:diffusion.dd} as a
perturbation of the DSA scheme, and define error
\begin{align*}
  \varepsilon &= f - h,
\end{align*}
which follows equation
\begin{align*}
  \left\{
    \begin{aligned}
      &\frac{-1}{3\Sigma} \, \varepsilon^{\prime\prime}(x) + (1-c)\,\Sigma\,\varepsilon(x) = 0, \\[0.5em]
      &\varepsilon^\prime(0) = \varepsilon^\prime(L) = 0, \\[0.5em]
      &\varepsilon(x_i) = \varepsilon_i, \qquad 1 \leqslant i \leqslant N-1.
    \end{aligned}
  \right.
\end{align*}

In the equation above, the value at subdomain interfaces is given by
\begin{align*}
  \varepsilon_i
  &= h(x_i) - f(x_i)
  = \frac{1}{2} \left[ g^-(x_i) + g^+(x_i) \right] - f(x_i)
  = \frac{1}{2} \left[ \delta^-(x_i) + \delta^+(x_i) \right].
\end{align*}
Equation~\eqref{eq:delta} yields
\begin{align}
  \delta^-(x_i) &= -f^\prime(x_{i-1}) \; \eNNl(l) - f^\prime(x_{i}) \; \eNNr(l), \notag\\
  \delta^+(x_i) &= -f^\prime(x_{i}) \; \eNNl(0) - f^\prime(x_{i+1}) \; \eNNr(0), \notag
\end{align}
and, noticing that terms evaluated at point $x_{i}$ vanish thanks to symmetry relation~\eqref{eq:enn.sym},
\begin{align}\label{eq:epsilon.interface}
  \varepsilon_i
  &= \frac{1}{2} \left[\vphantom\sum f^\prime(x_{i+1}) - f^\prime(x_{i-1})\right] \; \eNNl(l).
\end{align}

\bigskip

Following the same line of reasoning than for the Neumann diffusion problem, in
each internal subdomain ${\cal D}_i, \; 2\leqslant i\leqslant N-1$, error
$\varepsilon_{|{\cal D}_i}$ can be expressed as the linear combination
\begin{align}\label{eq:epsilon.interne}
  \varepsilon_{|{\cal D}_i} =
  \left[\varepsilon_{i-1} \; \eDDl + \varepsilon_{i} \; \eDDr \right] \circ t_i,
\end{align}
where $\eDDl$ and $\eDDr$ respectively measure errors stemming from not
knowing the value $h$ should take at the left and right extremities of the subdomain:
\begin{align*}
  \left\{
    \begin{aligned}
      &\frac{d^2\eDDl}{dx^2}(x) - \alpha^2 \; \eDDl(x) = 0, \\[0.5em]
      &\eDDl(0) = 1, \\[0.5em]
      &\eDDl(l) = 0,
    \end{aligned}
  \right.
  &\qquad\text{and}\qquad
  \left\{
    \begin{aligned}
      &\frac{d^2\eDDr}{dx^2}(x) - \alpha^2 \; \eDDr(x) = 0, \\[0.5em]
      &\eDDl(0) = 0, \\[0.5em]
      &\eDDl(l) = 1.
    \end{aligned}
  \right.
\end{align*}

As for the Neumann Diffusion problem, the solutions to these problem can be
analytically expressed, and are linked by symmetry relations:
\begin{align*}
  \eDDl(x) = -{{e^ {- \alpha\,x }\,\left(e^{2\,\alpha\,x}-e^{2\,\alpha\,l}\right)
 }\over{e^{2\,\alpha\,l}-1}}

  \quad\text{and}\quad
  \eDDr(x) = \eDDl(l-x).
\end{align*}

\bigskip

Boundary subdomains ${\cal D}_1$ and ${\cal D}_N$ must be handled specially,
since they have mixed boundary conditions: on one of their extremities, the
boundary condition is known exactly; the error only comes from not knowing the
exact boundary condition on the other extremity. We can write
\begin{align}
  \varepsilon_{|{\cal D}_1} = \varepsilon_1 \; \eNDr
  \quad\text{and}\quad
  \varepsilon_{|{\cal D}_N} = \varepsilon_{N-1} \; \eDNl \circ t_{N},
  \label{eq:epsilon.D1.Dn}
\end{align}
where $\eDNl$ and $\eNDr$ are given by
\begin{align*}
  \left\{
    \begin{aligned}
      &\frac{d^2\eDNl}{dx^2}(x) - \alpha^2 \; \eDNl(x) = 0, \\[0.5em]
      &\eDNl(0) = 1, \\[0.5em]
      &\frac{d\eDNl}{dx}(l) = 0,
    \end{aligned}
  \right.
  &\qquad\text{and}\qquad
  \left\{
    \begin{aligned}
      &\frac{d^2\eNDr}{dx^2}(x) - \alpha^2 \; \eNDr(x) = 0, \\[0.5em]
      &\frac{d\eNDr}{dx}(0) = 0, \\[0.5em]
      &\eNDr(l) = 1.
    \end{aligned}
  \right.
\end{align*}
As in previous cases, analytical and symmetric expressions can be found for
these terms:
\begin{align*}
  \eDNl(x) = {{e^ {- \alpha\,x }\,\left(e^{2\,\alpha\,x}+e^{2\,\alpha\,l}\right)
 }\over{e^{2\,\alpha\,l}+1}}

  \quad\text{and}\quad
  \eNDr(x) = \eDNl(l-x).
\end{align*}

\subsubsection{Flux correction}

After a PDSA iteration, the corrected scalar flux is given by
\begin{align*}
  \phiDD
  = \phiSI + h \notag
  = \phiSI + f + \varepsilon \notag
  = \rhoDSA \, \phi_0 + \varepsilon.
\end{align*}
Unlike in the standard DSA scheme, $\phi_0$ is not an eigenmode of the PDSA
scheme. It is therefore more difficult to express error evolutions from one
iteration to the next. It is however possible to state that
\begin{align}
  \frac{\left\Vert\phiDD\right\Vert_2}{\left\Vert\phi_0\right\Vert_2}
  &\leqslant \underbrace{\rhoDSA +
    \frac{\left\Vert\varepsilon\right\Vert_2}{\left\Vert\phi_0\right\Vert_2}}
    _{\rhoDDmax},
\label{eq:rhoDDmax}
\end{align}
where $\rhoDDmax$ denotes the upper bound of the amplification factor of the
whole PDSA scheme.

\medskip

Equations \eqref{eq:epsilon.interne}
and~\eqref{eq:epsilon.D1.Dn} yield
\begin{align*}
  \varepsilon
  &= \chi_1 \; \varepsilon_1 \; \eNDr
    + \sum_{i=2}^{N-1} \chi_i \; \left[\varepsilon_{i-1} \; \eDDl + \varepsilon_{i} \; \eDDr \right] \circ t_i
    + \chi_{N} \; \varepsilon_{N-1} \; \eDNl \circ t_{N},
\end{align*}
and
\begin{align*}
  \left\Vert\varepsilon\right\Vert_2^2
  &= \varepsilon_1^2 \; \left\Vert \eNDr \right\Vert_2^2
    + \sum_{i=2}^{N-1}\left\Vert\varepsilon_{i-1} \; \eDDl + \varepsilon_{i} \; \eDDr \right\Vert_2^2
    + \varepsilon_{N-1}^2 \; \left\Vert \eDNl\right\Vert_2^2 \notag\\
  &\leqslant \varepsilon_1^2 \; \left\Vert \eNDr \right\Vert_2^2
    + 2 \; \sum_{i=2}^{N-1} \left( \varepsilon_{i-1}^2 \; \left\Vert \eDDl\right\Vert_2^2 +
    \varepsilon_{i}^2 \; \left\Vert \eDDr \right\Vert_2^2 \right)
    + \varepsilon_{N-1}^2 \; \left\Vert \eDNl\right\Vert_2^2 \notag\\
  &\leqslant 5\;\left\Vert \eDDl \right\Vert_2^2\;\sum_{i=1}^{N} \varepsilon_i^2.
\end{align*}
The last inequality was obtained by noticing that
$\displaystyle
\left\Vert \eNDr \right\Vert_2^2
= \left\Vert \eDNl \right\Vert_2^2
\leqslant 3 \, \left\Vert \eDDl \right\Vert_2^2$
and
$\displaystyle
\left\Vert \eDDl \right\Vert_2^2
= \left\Vert \eDDr \right\Vert_2^2$.

\bigskip

Equation \eqref{eq:epsilon.interface} also allows to bound the error at interfaces
\begin{align*}
  \left\vert\varepsilon_i\right\vert
  &\leqslant \frac{1}{2}
    \left(\vphantom\sum \left\vert f^\prime(x_{i+1})\right\vert + \left\vert f^\prime(x_{i-1})\right\vert\right)
    \; \left\vert \eNNl(l)\right\vert \notag\\
  &\leqslant \left\vert \eNNl(l)\right\vert \; \sup_{x} \left\vert f^\prime(x)\right\vert \notag\\
  &\leqslant \left\vert\rho_d\right\vert \; \left\vert \eNNl(l)\right\vert \; \sup_{x} \left\vert \phi_0^\prime(x)\right\vert \notag\\
  &\leqslant \left\vert\rho_d\right\vert \; \frac{k\,\pi}{N\,l} \; \left\vert \eNNl(l)\right\vert,\notag
\end{align*}
so that
\begin{align*}
\varepsilon_i^2
  &\leqslant \rho_d^2 \; \left(\frac{k\,\pi}{N\,l}\right)^2 \; \left(\eNNl(l)\right)^2.
\end{align*}

Combining previous results yields the following global bound:
\begin{align*}
  \left\Vert\varepsilon\right\Vert_2
  &\leqslant \sqrt{5\;N}\;\rho_d\;\frac{k\,\pi}{N\,l}\;
    \left|\eNNl(l)\right|\;\Vert \eDDl \Vert_2.
\end{align*}

\begin{figure}[t]
  \begin{minipage}{0.47\linewidth}
    \includegraphics[width=\linewidth]{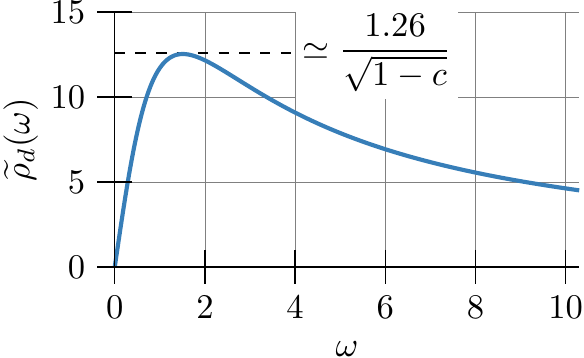}
    \caption{Evolution of factor $\rhoDD$ with frequency $\omega$, for
      $c=0.99$.}
    \label{fig:rhoDD}
  \end{minipage}
  \hfill
  \begin{minipage}{0.47\linewidth}
    \includegraphics[width=\linewidth]{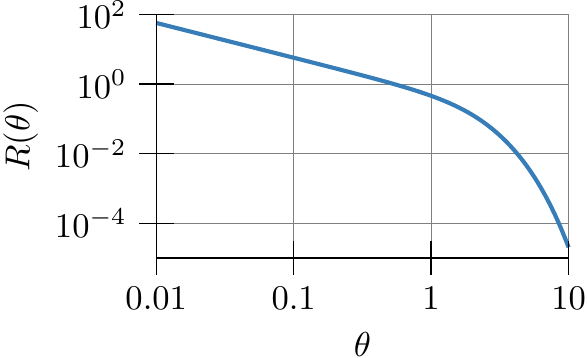}
    \caption{Evolution of factor $R$ with optical thickness $\theta$.}
    \label{fig:R}
  \end{minipage}
\end{figure}

Noticing that, as soon as $k\neq 0$,
$\left\Vert\phi_0\right\Vert_2 = \sqrt{\frac{L}{2}} = \sqrt{\frac{N\,l}{2}}$,
it follows that
\begin{align}
  \frac{\left\Vert\varepsilon\right\Vert_2}{\left\Vert\phi_0\right\Vert_2}
  &\leqslant \sqrt{\frac{2}{N\;l}} \; \sqrt{5\;N}\;\rho_d\;\frac{k\,\pi}{N\,l}\;
    \left|\eNNl(l)\right|\;\Vert \eDDl \Vert_2\notag\\
  &\leqslant \underbrace{\sqrt{\frac{10}{3\,(1-c)}} \;\rho_d\;\omega}_{\rhoDD(\omega)}
    \;
    \underbrace{\vphantom{\sqrt\frac{1}{(2)}}\frac{\alpha\;\left|\eNNl(l)\right|\;\Vert \eDDl \Vert_2}{\sqrt{l}}}_{R} 
  \label{eq:error}
\end{align}
It should be mentioned that the first part of this expression, denoted by
$\rhoDD$, only depends on the scattering ratio~$c$ and the
frequency~$\omega$. As shown by an asymptotic development and illustrated in figure~\ref{fig:rhoDD}, in the asymptotic limit
when $c\rightarrow 1$, the maximum value of $\rhoDD$ is
approximately given by~$\frac{1.26}{\sqrt{1-c}}$.

On the other hand, the second part of expression~\eqref{eq:error}, denoted by
$R$, can be expressed as
\begin{align*}
  R(\theta) &= \sqrt{{{2\,e^{6\,\theta}-8\,\theta\,e^{4\,\theta}-2\,e^{2\,\theta}
 }\over{\theta\,e^{8\,\theta}-4\,\theta\,e^{6\,\theta}+6\,\theta\,e^{
 4\,\theta}-4\,\theta\,e^{2\,\theta}+\theta}}}
,
\end{align*}
where we defined quantity
\begin{align*}
  \theta &= \alpha\,l = \sqrt{3\;(1-c)} \; \frac{\Sigma \; L}{N},
\end{align*}
which is a dimensionless parameter depending only on physical properties
associated to the problem, and characterizes the optical thickness of a
subdomain. Figure~\ref{fig:R} presents the variation of factor $R$ with optical
thickness~$\theta$. As shown by asymptotic developments for small and large
optical thicknesses, $R$ is not bounded for small optical thicknesses, but
converges extremely rapidly towards~0 when the optical thickness of subdomains
increases:
\begin{align*}
  R(\theta) \mathop{\quad\sim\quad}_{\theta\rightarrow 0} {{1}\over{\sqrt{3}\,\theta}}

  \qquad\text{and}\qquad
  R(\theta) &\mathop{\quad\sim\quad}_{\theta\rightarrow \infty} {{\sqrt{2}\,e^ {- \theta }}\over{\sqrt{\theta}}}
.
\end{align*}

\subsubsection{Convergence}
\label{sec:fourier.pdsa.conv}

\newcommand{\lc}{\underline{l}}

As a conclusion, for any set of cross sections $\Sigma$ and $\sigs$, there exists
a critical subdomain size $\lc$ such that
\begin{align}
  \forall l \geqslant \lc, \qquad
  \frac{\left\Vert\varepsilon\right\Vert_2}{\left\Vert\phi_0\right\Vert_2}
  &< 1 - \rhoDSA, \notag\\
  \intertext{so that, from equation~\eqref{eq:rhoDDmax},}
  \frac{\left\Vert\phiDD\right\Vert_2}{\left\Vert\phi_0\right\Vert_2}
  &\leqslant \rhoDDmax < 1,  \label{eq:critical.size}
\end{align}
and the PDSA scheme can accelerate the convergence of source
iterations. Moreover, as the subdomain size $l$ increases above the critical
size, the efficiency of the PDSA scheme very rapidly converges to that of the
standard DSA scheme:
\begin{align*}
  \frac{\left\Vert\phiDD\right\Vert_2}{\left\Vert\phi_0\right\Vert_2}
  \; \xrightarrow[l\to\infty]{} \; \rhoDSA.
\end{align*}

Conversely, since $\rhoDD\xrightarrow[c\to 0]{} 0$, for any domain of fixed
optical thickness \mbox{$\tau = \Sigma\,L$}, there exists a critical scattering
ratio~$\bar c$ under which the PDSA scheme converges:
\begin{align*}
  \forall c \leqslant \bar c,
  \qquad   \frac{\left\Vert\phiDD\right\Vert_2}{\left\Vert\phi_0\right\Vert_2}
  &\leqslant \rhoDDmax < 1.
\end{align*}

\bigskip

In practice, this limits the use of the PDSA scheme to cases which are optically
thick enough for condition~\eqref{eq:critical.size} to apply for the whole
geometrical domain. In such cases, the condition also limits the maximal number
of subdomains which can be used.

\subsubsection{Special Case: Two-Subdomain Partition}
\label{sec:fourier.pdsa.2dom}

In the special case where the domain is partitioned in two subdomains, the first
Neumann diffusion step in the PDSA scheme yields, from
equation~\eqref{eq:epsilon.interface} and boundary conditions from
problem~\eqref{eq:diffusion.glob}:
\begin{align*}
  \varepsilon_1
  &= \frac{1}{2} \left[\vphantom\sum f^\prime(0) - f^\prime(L)\right] \; \eNNl(l) = 0.
\end{align*}
It follows that the second PDSA step, Dirichlet diffusion, yields the solution
to the global DSA problem: $h = f$.  In this case, the PDSA scheme is thus
equivalent to a global DSA scheme.

\section{Numerical results}
\label{sec:num}

In order to assess the validity of the above theory, we present in this section
some numerical results.

These results were obtained using a very simple code, developed in Julia. We
consider the time-independent, one-group Boltzmann equation with isotropic
scattering, set in an homogeneous 1D slab geometry over the spatial
domain~$[0,L]$. In order to model a full core, we set void boundary conditions
with no incoming flux:
\begin{align*}
  \left\{
    \begin{aligned}
      &\mu \frac{\partial\psi}{\partial x}(x,\mu) + \Sigma\,\psi(x,\mu)
      = \frac{\sigs}{2}\,\int_{-1}^1 \psi(x, \mu') \; d\mu' + Q(x), \\[0.5em]
      &\psi(0,\mu) = 0 \qquad \forall\mu>0,\\
      &\psi(L,\mu) = 0 \qquad \forall\mu<0.\\
    \end{aligned}
  \right.
\end{align*}

The solver uses the discrete-ordinates method to handle the angular dependency
of the solution. The transport equation is spatially discretized using a
standard diamond-differencing (DD) scheme. The diffusion equations used in the
(P)DSA schemes are discretized using a $P_1$ finite-element method.

\medskip

In the following, we will set a unit-length domain~(\mbox{$L=1$}) and a linear
source~(\mbox{$Q(x)=x$}). The cases studied will vary only with respect to the
material used in the geometry, which can be entirely characterized by its total
and scattering cross-sections $\Sigma$ and $\Sigma_s$. Equivalently, the problem
may be characterized by its total optical thickness $\tau=\Sigma\,L$ and its
absorption ratio \mbox{$\epsilon = 1-c = 1-\frac{\Sigma_s}{\Sigma}$}.

From the bounds discussed above, one may expect the PDSA to converge easily for
large values of~$\tau$ and~$\epsilon$.

\subsection{Fourier analysis}

\begin{figure}[tb]
  \begin{center}
    \includegraphics[width=\linewidth]{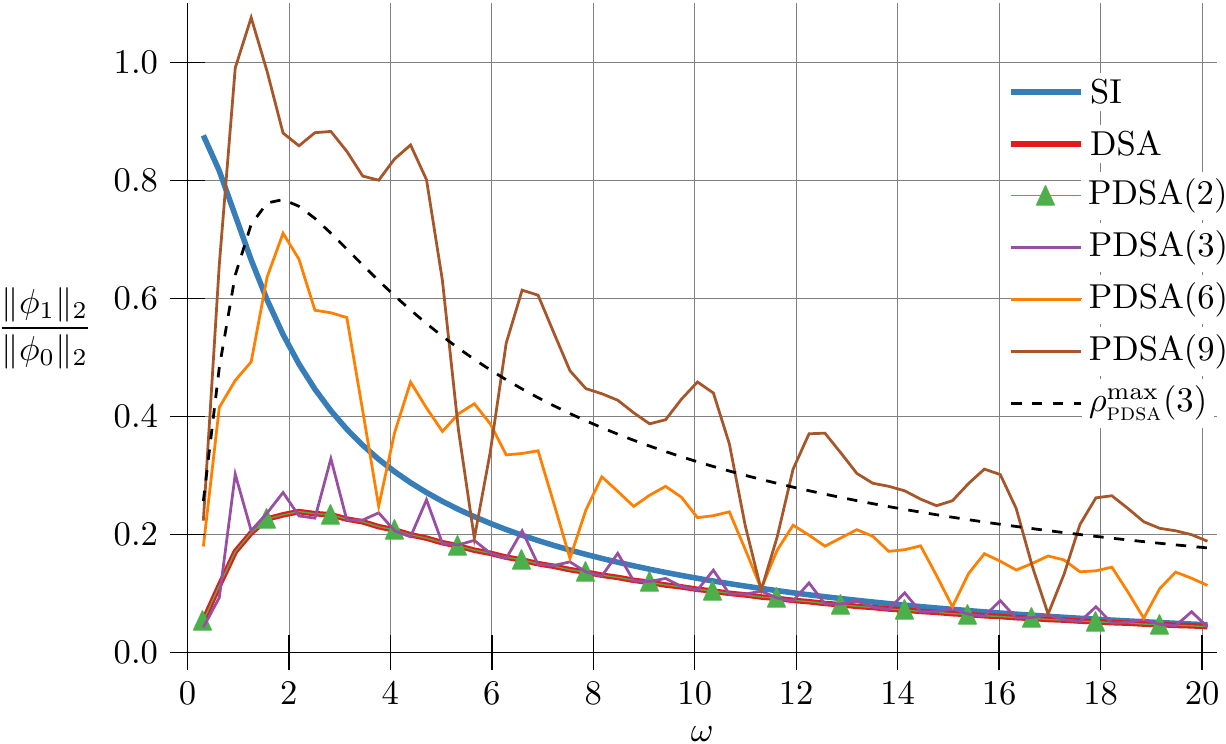}
  \end{center}
  \caption{Measured amplification factors of various acceleration schemes, for
    the case where $\tau = 10$ and $\epsilon = 0.1$ (case~A in
    table~\ref{tab:source}).}
  \label{fig:fourier1}
\end{figure}

Setting an initial flux of the form given by~\eqref{eq:phi0} and performing an
iteration, one can perform a numerical Fourier analysis of the different
schemes.

The results of such an analysis are presented in figure~\ref{fig:fourier1}, in
the case where \mbox{$\tau=10$} and~$\epsilon=0.1$. Unsurprisingly, the Source
Iterations and DSA schemes behave similarly to figure~\ref{fig:rho}. The
behavior of the PDSA scheme is presented for different numbers of subdomains. As
noted in paragraph~\ref{sec:fourier.pdsa.2dom}, the PDSA scheme with two
subdomains is exactly equivalent to the standard DSA scheme. Then, as the
number of subdomains increases, larger and larger perturbations start to appear
until the amplification factor exceeds~1 for 9 subdomains.

The dashed black line in figure~\ref{fig:fourier1} represents the theoretical
bound on the amplification factor, as obtained using
eqs.~\eqref{eq:rhoDDmax}--\eqref{eq:error} in the case of 3~subdomains. It
appears that this value effectively bounds the measured amplification factor,
but is not very sharp.

\begin{figure}[tb]
  \begin{center}
    \includegraphics[width=\linewidth]{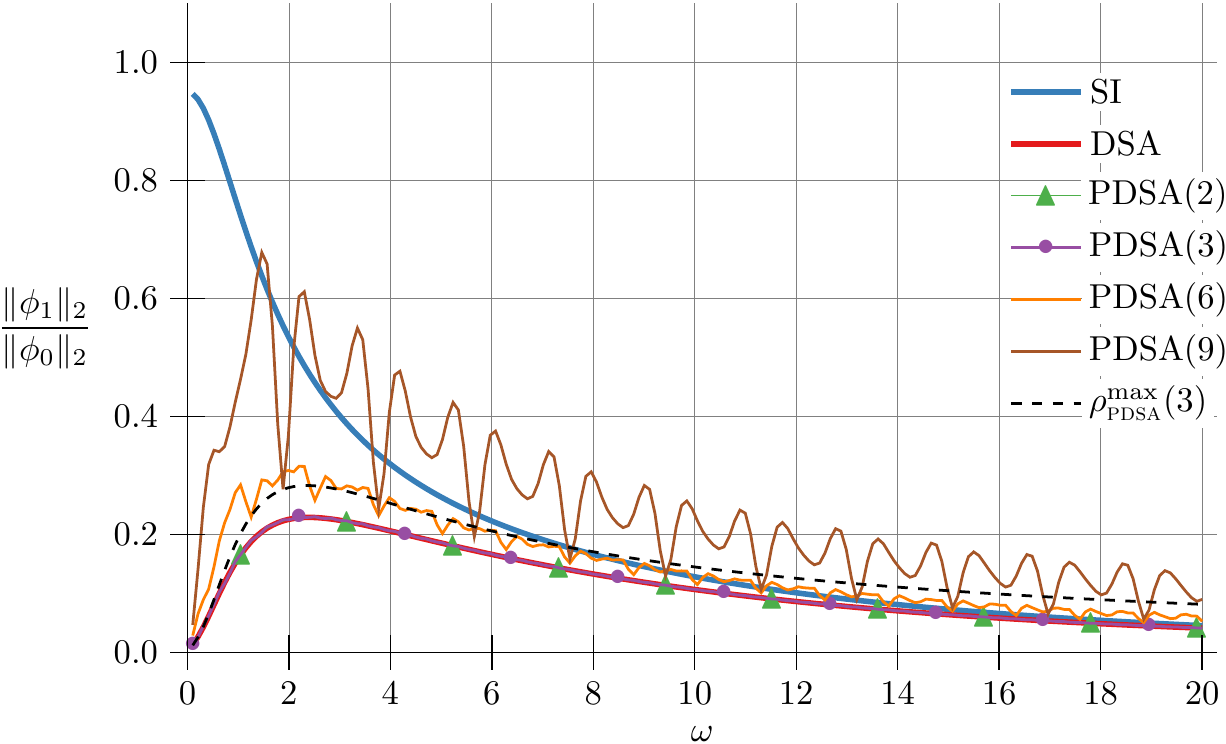}
  \end{center}
  \caption{Measured amplification factors of various acceleration schemes, for
    $\tau = 30$ and $\epsilon = 0.05$ (case~B in table~\ref{tab:source}).}
  \label{fig:fourier2}
\end{figure}

However, such a problem being neither very optically thick nor very diffusive,
it is not representative of the cases where PDSA would be applied in practice
for PWR calculations. In order to show a tendency when the optical thickness
increases, figure~\ref{fig:fourier2} presents the same analysis for
\mbox{$\tau = 30$} and \mbox{$\epsilon=0.05$}. In this case, both 2-domain and
3-domain PDSA are indistinguishable from the standard DSA, and amplification
factors for other numbers of subdomains are reduced as expected. The theoretical
bound for PDSA(3) is still over-evaluated, but stays in more acceptable limits.

\subsection{Number of iterations}

The practical interest of the PDSA scheme can be assessed in terms of reduction of the number of
iterations. Table~\ref{tab:source} presents a comparison of the acceleration
schemes on different problems. An ``X'' marks settings in which the PDSA scheme
does not converge.

The first two rows of the table (cases A and B) correspond to the two cases used
for the Fourier analysis in the previous paragraph. In case A, we can see that,
as expected, 9-domain PDSA does not converge in the first case. However,
although figure~\ref{fig:fourier1} showed rather degraded amplification factors
for the 6-domain PDSA, its iteration count is in practice not higher than for
the standard DSA. Similar results occur for case B, in which all PDSA schemes
exhibit no degradation of efficiency with respect to the standard DSA. This is
in contrast to figure~\ref{fig:fourier2}, which evidenced a degradation of the
amplification factor for 9-domain PDSA.

Cases C--E demonstrate the behaviour of the iterations count as $\epsilon$
decreases. Unsurprisingly, the number of source iterations increases with the
scattering ratio. This is in contrast with the rather stable DSA iterations
count. The PDSA schemes behave almost identically to DSA, until they reach a
point where the number of iterations starts increasing. The scheme stops
converging soon after this point.

\begin{table}[tb]
  \centering
  \begin{tabular}{ccccccccc}
    \toprule
     & $\tau$ & $\epsilon$ & SI & DSA & PDSA(3) & PDSA(4) & PDSA(6) & PDSA(9)\\
\midrule
A & 10 & 0.100 & 144 & 24 & 24 & 24 & 24 & X\\
B & 30 & 0.050 & 342 & 27 & 27 & 27 & 27 & 27\\
C & 30 & 0.010 & 1399 & 34 & 34 & 34 & 56 & X\\
D & 30 & 0.005 & 2248 & 35 & 36 & 36 & X & X\\
E & 30 & 0.001 & 4351 & 38 & X & X & X & X\\

    \bottomrule
  \end{tabular}
  
  \caption{Iterations count of the various schemes for several cases.}
  \label{tab:source}
\end{table}

\begin{figure}[tb]
  \begin{center}
    \includegraphics[width=\linewidth]{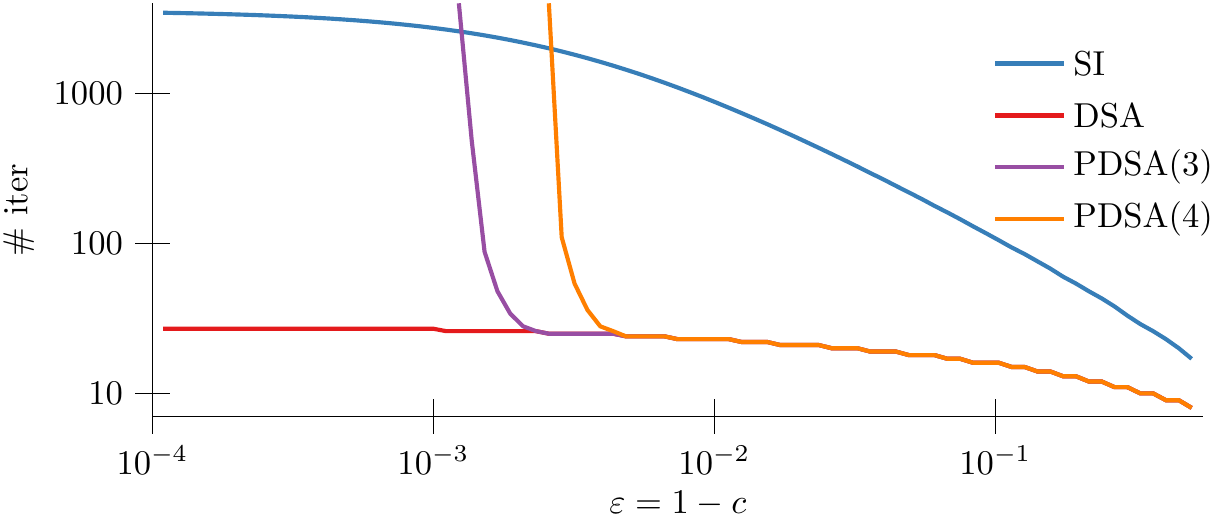}
  \end{center}
  \caption{Iteration count as a function of the scattering ratio, for $\tau = 30$.}
  \label{fig:iter}
\end{figure}

This is more clearly shown on figure~\ref{fig:iter}, which presents the
variation of the iteration count with the scattering ratio. Reading the figure
from right to left: as the scattering ratio increases, the number of source
iterations increases. At the same time, the DSA iterations count stays more
stable. PDSA behaves identically to DSA, until the scattering ratio approaches a
critical value, at which its performances degrade very rapidly. This evidences
the existence of a critical scattering ratio $\bar c$, as mentioned in
section~\ref{sec:fourier.pdsa.conv}.

\medskip

The main conclusion to draw from this study is that, when it converges, the PDSA
scheme almost always exhibits the same performance as the standard DSA. The
following part discusses the conditions under which PDSA does converge.

\subsection{Convergence region}

\begin{figure}[tb]
  \begin{center}
    \includegraphics{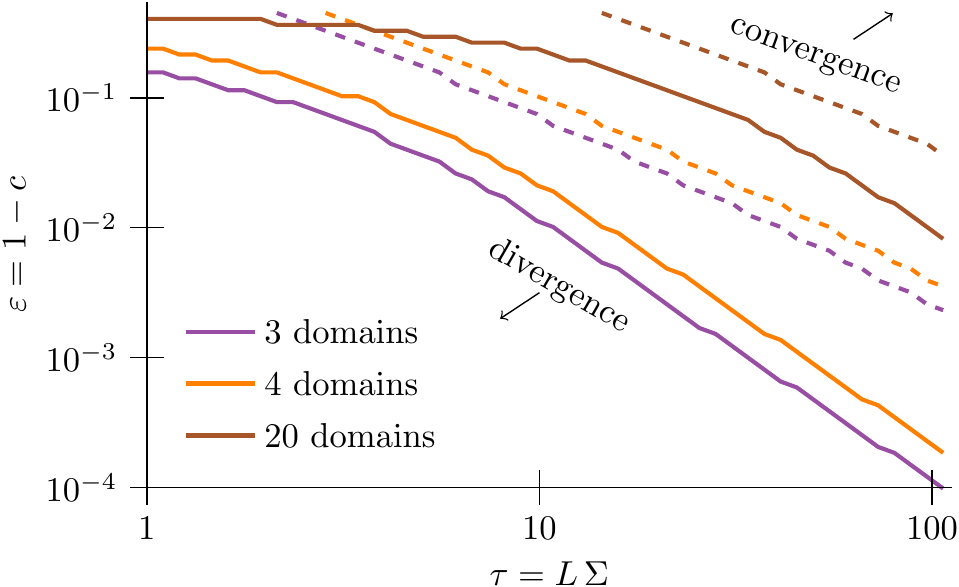}
  \end{center}
  \caption{Boundary of the domain of convergence of PDSA: the scheme converges
    only for parameters which are above the curves. Theoretical limits are
    indicated by dashed lines, while solid lines indicate the experimentally
    measured limits.}
  \label{fig:convergence}
\end{figure}

Figure~\ref{fig:convergence} presents the convergence region of the PDSA
scheme. For low scattering ratios and high optical lengths (in the top right
part of the figure), PDSA converges. Then, as the scattering ratio increases
above the critical value~$\bar c$, the scheme leaves the convergence region. On
figure~\ref{fig:convergence}, dashed lines present the theoretical critical
scattering ratio. That is, the dashed lines are level curves for which
$\rhoDDmax = 1$. On the other hand, solid lines present the critical point at
which the scheme is observed to start diverging in practice.

First, it is interesting to note that the theoretical value always bounds the
practical one. In other words, for a given number of subdomains, the dashed line
is always above the solid one. The overestimation of~$\rhoDDmax$, observed in
figures~\ref{fig:fourier1} and~\ref{fig:fourier2}, manifests itself as a gap
between the theoretical and observed values. In practice, theoretical bounds can
help ensuring that the PDSA scheme will converge when $\rhoDDmax<1$. However, if
the theoretical bound goes above unity, a practical test should still be
conducted, as the PDSA might still very well converge. This is especially true
in the limit of large optical thicknesses, where the overestimation
of~$\rhoDDmax$ seems to increase.

\section{Conclusions}
\label{sec:concl}

We presented in this paper a piecewise Diffusion Synthetic Acceleration Scheme
(PDSA), which is specifically designed to be straightforwardly used in parallel
contexts. The implementation of PDSA only requires having a standard neutron
diffusion solver whose discretization is consistent with that of the neutron
transport solver. In practice, and as explained in~\cite{moustafa20xx}, starting
from an initially sequential DSA-accelerated transport code, one only needs to
take care of the parallelization of the transport solver; the parallel
acceleration scheme comes at no practical development cost.

We showed that, although the PDSA scheme only approximates DSA, it converges for
a class of problems which are optically thick enough. For this class of
problems, we also showed that PDSA is in practice as efficient as standard DSA,
in terms of the number of iterations. We presented an indicator, coming from 1D
geometries but computable for any kind of 3D problem, allowing to estimate
\textit{a priori} if the problem at hand is optically thick enough for PDSA to
converge.

This indicator is the main shortcoming of this work. Simple 1D experiments in
this work show that PDSA performs in practice much better than the indicators
would predict. The work presented in~\cite{moustafa20xx} draws similar
conclusions for more complex, 3D, industrial calculations. In practice, the
theoretical indicator presented here can be used to guarantee that the method
will converge, but no practical conclusion can be drawn as to the divergence of
the scheme. This might be because the bounds derived here are not tight enough
to yield the sharp estimators that one would like to have in practice. Also, as
source iterations advance, we might expect the DSA correction to be smoother and
smoother, and the gradient of the correction to be closer and closer to
zero. This phenomenon has not been accounted for here, although it could help
limiting the error made in the first PDSA step. This should be the topic of
further analyses and work.

\bibliography{theorie}
\bibliographystyle{elsarticle-num}

\end{document}